# Probing dense QCD matter in the laboratory – The CBM experiment at FAIR

P. Senger[1,2] for the CBM Collaboration

[1] Facility for Antiproton and Ion Research, Darmstadt, Germany

[2] National Research Nuclear University MEPhI, Moscow, Russia

E-mail: p.senger@gsi.de

**Abstract**

The "Facility for Antiproton and Ion Research" (FAIR) in Darmstadt will provide unique research opportunities for the investigation of fundamental open questions related to nuclear physics and astrophysics, including the exploration of QCD matter under extreme conditions, which governs the structure and dynamics of cosmic objects and phenomena like neutron stars, supernova explosions, and neutron star mergers. The physics program of the Compressed Baryonic Matter (CBM) experiment is devoted to the production and investigation of dense nuclear matter, with a focus on the high-density equation-of-state (EOS), and signatures for new phases of dense QCD matter. According to the present schedule, the CBM experiment will receive the first beams from the FAIR accelerators in 2025. This article reviews promising observables, outlines the CBM detector system, and presents results of physics performance studies.

1. **The future Facility for Antiproton and Ion Research (FAIR)**

The Facility for Antiproton and Ion Research (FAIR) will be the accelerator-based flagship research facility in Europe for the coming decades. FAIR will open up unprecedented research opportunities in nuclear and hadron physics including nuclear astrophysics, in atomic and plasma physics, and in applied research such as material science and radiation biophysics [1]. The layout of FAIR is shown in Figure 1. The synchrotron SIS100 will provide high-intensity primary beams like protons up to kinetic energies of 29 GeV, and nuclei with energies up to 15A GeV. Intense beams of rare isotopes will be produced in nuclear collisions, selected by the Superconducting Fragment Separator (SFRS), and delivered for further investigation to the experimental setups build by the NUSTAR collaboration (**Nu**clear **St**ructure, **A**strophysics and **R**eactions). The research program at the SFRS will include the measurement of the masses and lifetimes of very neutron rich or neutron deficient nuclei, in order to explore the paths of nucleosynthesis, and hence, shed light on the origin of elements in the universe. The High-Energy Storage Ring (HESR) will accelerate and cool intense secondary beams of antiprotons to be used for experiments on hadron physics, including charmonium spectroscopy and the search for gluonic excitations, with a detector system realised by the PANDA collaboration (Anti**P**roton **An**nihilation at **Da**rmstadt). The atomic physics research program includes test of quantum electrodynamics in the range of strong fields, and spectroscopy of anti-hydrogen stored in a low-energy ring. The experiments on plasma physics will study matter under conditions like in the interior of stars and planets. Biophysics experiments with high-intensity beams will pursue the development of particle therapy, and will study space radiation effects in cooperation with the European Space Agency (ESA). The Compressed Baryonic Matter (CBM) detector is designed for the investigation of high-energy nucleus-nucleus collisions, where matter densities are created like in the core of neutron stars.

Civil construction of the accelerator ring and the CBM cave is advancing according to schedule, and the call for tender for the construction of the buildings for the SFRS and the other experiments is ongoing. Installation and commissioning of the experiments is planned for 2022-2024, and FAIR will be operational in 2025.

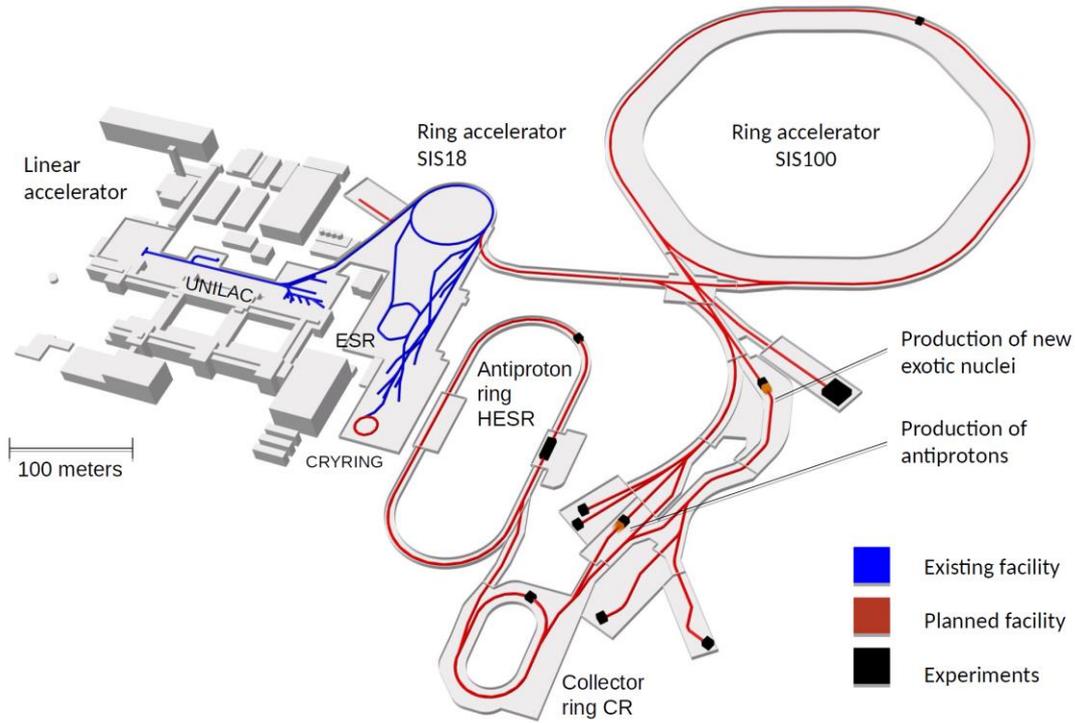

Fig. 1: Layout of the future Facility for Antiproton and Ion Research [2].

The beam energies and ion species provided by the SIS100 synchrotron at FAIR are ideally suited for the production of high-density baryonic matter in collisions between heavy nuclei. According to microscopic transport calculations and hydro-dynamical models, baryon densities of more than 5 times the density of an atomic nucleus are produced in central Au+Au collision at 5A GeV [3]. At higher densities, which might be reached at beam energies of 10A GeV, the nucleons are expected to percolate and melt into quarks and gluons [4]. The following article focuses on future high-precision measurements of observables from heavy-ion collision experiments, which are predicted to probe the high-density nuclear equation-of-state, and are expected to indicate new phases of QCD matter at large net-baryon density.

2. *The high-density nuclear equation-of-state*

The EOS describes the relation between density, pressure, volume, temperature, energy, and isospin asymmetry. For a constant temperature, the pressure can be written as

$$P = \rho^2 \, \delta(E/A)/\delta\rho$$

where P represents the pressure, $\rho$ the density, and E/A the energy per nucleon, which is defined as

$$E/A(\rho,\delta) = E/A(\rho,0) + E_{sym}(\rho)\cdot\delta^2 + O(\delta^4)$$

with the symmetry energy $E_{sym}$ and the asymmetry parameter $\delta = (\rho_n - \rho_p)/\rho$. The EOS for symmetric matter has a minimum value at saturation density $\rho_0$, and a curvature $K_{nm} = 9\rho^2 \, \delta^2(E/A)/\delta\rho^2$ with $K_{nm}$ the nuclear incompressibility of symmetric matter. In measurements of giant monopole resonances of heavy nuclei, i.e. at saturation density, the nuclear incompressibility has been found to be $K_{nm}(\rho_0) = 230\pm10$ MeV [5]. Various EOS calculated for symmetric nuclear are shown in figure 2 [6].

In experiments at GSI, the EOS of symmetric matter has been studied up to densities of about 2 $\rho_0$. The KaoS collaboration measured subthreshold $K^+$ production in a very heavy and a light collision system at different beam energies [7]. According to microscopic transport calculations, in heavy-ion collisions at kinetic beam energies below 1.6A GeV, i.e. below the $K^+$ production threshold in nucleon-nucleon collisions, $K^+$ mesons are produced in secondary collisions of pions and nucleons. These processes are enhanced at higher densities, and, therefore, are sensitive to the EOS. The kaon data could only be reproduced by RQMD and IQMD model calculations when assuming a soft EOS ($K_{nm}$=200 MeV), and taking into account in-medium effects [8,9]. The elliptic flow of protons, deuterons, tritons and $^3$He has been measured by the FOPI collaboration in Au+Au collisions at energies from 0.4A to 1.5A GeV. IQMD transport calculations could only reproduce the experimental results using a soft EOS ($K_{nm}$=190±30 MeV) and momentum-dependent interactions [10]. These results rule out the Skyrme type EOS (blue-dashed line, K=380) in figure 2.

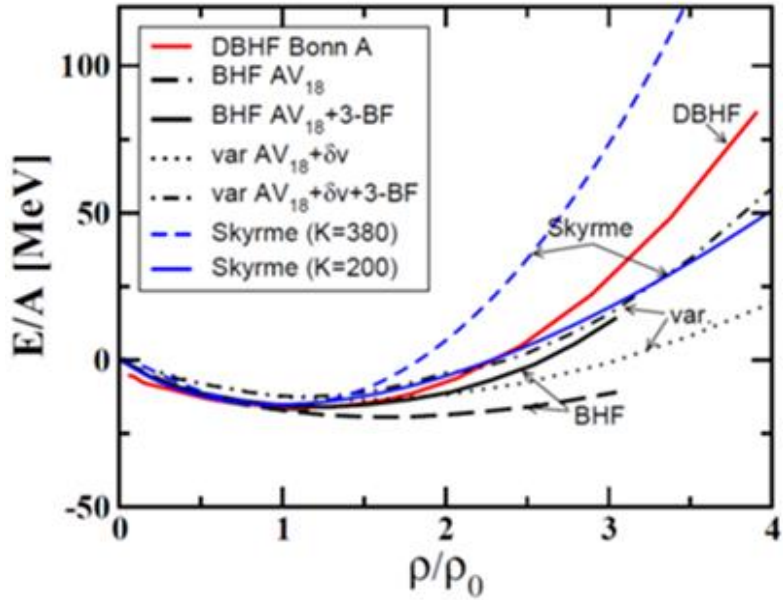

Fig.2: The isospin-symmetric nuclear matter EOS based on Skyrme forces for different incompressibility compared to the predictions from microscopic ab initio calculations [6].

The FOPI and ASY-EOS collaborations at GSI have also studied the symmetry energy at baryon densities above saturation density in Au+Au collisions 400A MeV by measuring the elliptic flow of neutrons and protons [11,12]. By comparing the FOPI data to the results of UrQMD transport calculations, a value of about $E_{sym}$= 60 ±10 MeV was found for 2 $\rho_0$ [11]. From the ASY-EOS data a value of about $E_{sym}$= 55±5 MeV at 2 $\rho_0$ was extracted [12]. These results rule out the very soft and very hard EOS depicted in figure 3.

In order to study the high-density EOS, which is relevant for the understanding of the most massive neutron stars, densities well above twice saturation densities should be produced and investigated in the laboratory. This is illustrated in figure 4, where the mass of neutron stars is plotted versus their central density for various equations-of-states [14]. The recent relativistic Shapiro delay measurement of an extremely massive millisecond pulsar found a mass of 2.14 + 0.10 − 0.09 solar masses [15], which corresponds, according to figure 4, to a central density between 4 and 7 $\rho_0$ depending on the EOS.

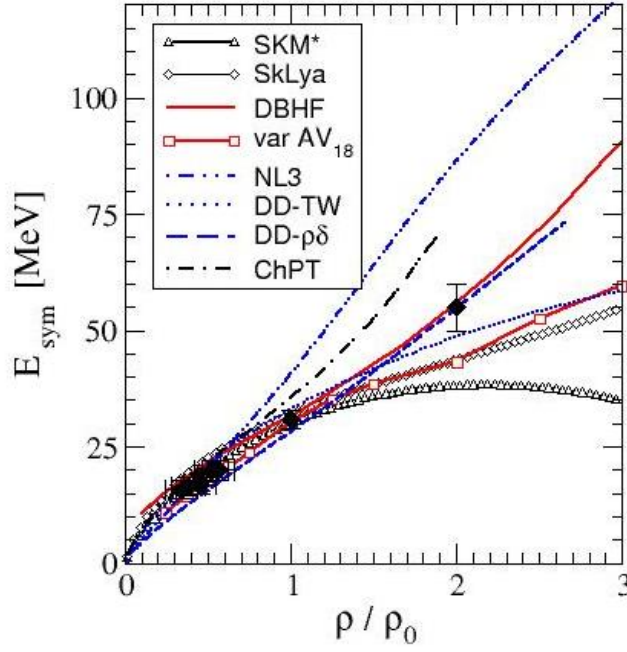

Fig.3: Symmetry energy for different EOS as function of density. Taken from [6]. Values of $E_{sym}$ below saturation density have been obtained in experiments with isospin-asymmetric nuclei [13]. The data point at 2 $\rho_0$ is extracted from measurements of the elliptic flow of neutrons and protons [12].

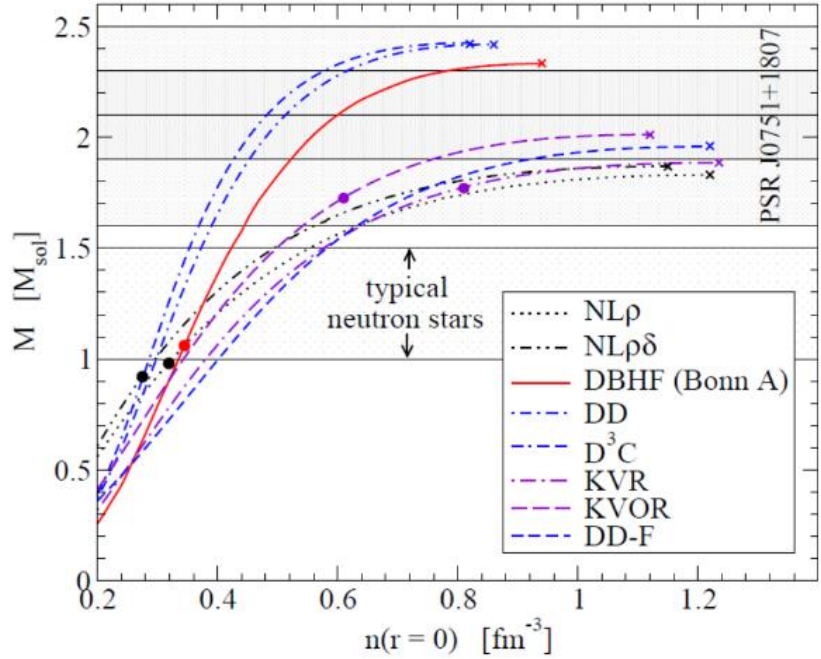

Fig. 4: Mass versus central density of neutron stars calculated from the Tolman–Oppenheimer–Volkoff (TOF) equation using different equations-of-state (EOS) [14].

Pioneering experiments on the high-density equation-of-state of symmetric nuclear matter have been performed at the AGS in Brookhaven by investigating the collective flow of protons in Au+Au collisions at kinetic energies from 2A GeV to 11A GeV [16]. The collective flow of hadrons is generated by the pressure gradient in the reaction zone, and, hence, is sensitive to the EOS. According to relativistic transport model calculations, the directed flow data could be explained best by a soft EOS ($K_{nm}$ = 210 MeV), whereas the elliptic flow data are better reproduced by the assumption of a stiff EOS ($K_{nm}$ = 300 MeV) [17].

In conclusion, the flow data measured in heavy-ion collisions at beam energies above 2A GeV, i.e. for baryon densities above about twice saturation density, are still consistent with values for $K_{nm}$ between 210 MeV and 300 MeV, and only rule out extreme values for the nuclear incompressibility.

The research program of the future Compressed Baryonic Matter experiment at FAIR includes precise measurements of the collective flow of protons, light fragments, and (multi-) strange particles for energies between 2A GeV and 11A GeV, and, therefore, will substantially improve the data situation. Moreover, the yield of multi-strange (anti-) hyperons, produced at beam energies close to their production threshold, is expected to be also sensitive to the high-density EOS of symmetric matter. The reason is, that in this case the $\Xi$ and $\Omega$ hyperons are created in multiple collisions involving kaons, lambda and sigma hyperons [18, 19]. These processes occur more frequently if the fireball density is high, which in turn depends on the stiffness of the EOS. The kaon production experiments at GSI have demonstrated, that the sensitivity of the kaon yield to the EOS increases with decreasing beam energy, i.e. if the beam energy is well below the kaon production threshold.

According to preliminary results of PHQMD transport model calculations, the yields of $\Xi$ and $\Omega$ hyperons created in Au+Au collisions at FAIR energies are very sensitive to the EOS [20]. The yield of multi-strange hyperons are significantly higher for a soft than for a hard EOS. For a given beam energy, this effect increases with increasing hyperon masses, because their threshold is higher, and their production depends strongly on the energy accumulation via multiple collisions in the dense medium. Therefore, the CBM experiment at FAIR will measure the excitation functions of multi-strange (anti-) hyperons in Au+Au collisions, and, for reference, in C+C collisions. These investigations devoted to the high-density EOS of nuclear matter are an important part of the CBM research program with a substantial discovery potential.

In addition to the EOS for symmetric matter, it is important to study the symmetry energy at high densities in order to contribute to our understanding of neutron stars. One possibility is to extend the measurements of the elliptic flow of neutrons and charged particles to higher beam energies, i.e. to higher densities. Another option is to measure particles of opposite isospin, i.e. with $I_3 = \pm 1$, which can be related to the densities of neutrons and protons. For example, transport models predict the $\pi^-/\pi^+$ ratio to be sensitive to the symmetry energy if measured at the threshold for pion production. The analysis of pion yields in Au+Au collisions at 400A MeV performed by the FOPI collaboration have been found not to be conclusive, as the beam energy is well above threshold, and the sensitivity of the $\pi^-/\pi^+$ ratio to the symmetry energy decreases with increasing beam energy [21]. Furthermore, the $\pi^-/\pi^+$ ratio is also effected by the unknown in-medium potential of the $\Delta(1232)$ resonance [22]. At FAIR energies, it would be worthwhile to explore the sensitivity of the $\Sigma^-/\Sigma^+$ ratio to the high-density symmetry energy. As one of the decay products of $\Sigma$ hyperons is always a neutral particle, the reconstruction of $\Sigma$ hyperons is based on the missing mass method using the tracking detectors of the CBM experiment.

*3. The degrees-of-freedom of QCD matter at high densities*

When exploring the EOS at neutron star core densities, the question arises, what the degrees of freedom of matter under such conditions are. For very low net-baryon densities, the fundamental theory of strong interaction, Quantum Chromo Dynamics (QCD), predicts a smooth crossover transition from hadronic matter to a deconfined phase consisting of quarks and gluons at a pseudo-critical temperature of about 155 MeV [23,24]. This is indicated in figure 5, which illustrates the QCD phase diagram as function of temperature, baryon-chemical and isospin-chemical potentials [25]. Up to now, the validity of QCD is still limited to the region of relatively low baryon chemical potentials, and, hence, leaves room for effective models, which predict structures like a first-order phase transition with a critical endpoint, or exotic phases such as quarkyonic matter [26,27]. In addition to these conjectured landmarks, also the regions of cosmic matter like neutron stars and neutron star mergers are displayed in figure 5.

It is commonly expected, that the nucleons percolate and melt into quarks and gluons at baryon densities above about 4-5 times saturation density. However, whether this phase transition is of first order with a critical endpoint as suggested in figure 5, or whether it proceeds smoothly by a crossover, is still a matter of debate. For example, following the concept of quark-hadron continuity, it is predicted that quark degrees of freedom emerge gradually with increasing density and partial restoration of chiral symmetry [4]. In this case, there would be no first order transition at high densities and low temperatures.

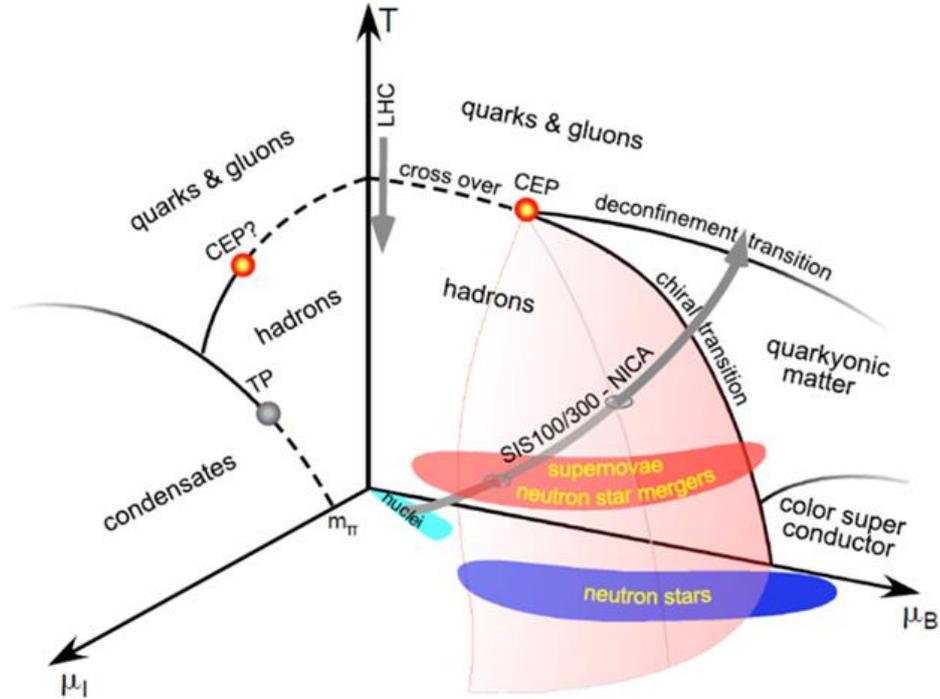

Fig. 5: Illustration of the QCD phase diagram as a function of temperature, baryon-chemical and isospin-chemical potential, including possible structures and astrophysical sites. Taken from [25].

Densities up to 4 $\rho_0$ are expected to be reached also in neutron star mergers as illustrated in figure 6, which depicts a snapshot of the equatorial plane illustrating the evolution of a neutron star merger with a total mass of 2.8 solar masses calculated with a Chiral Mean Field model [28]. The left part of the plot displays the temperature T, while the right part presents the quark fraction $Y_{quark}$. The green lines represent contours of constant baryon density in units of the nuclear saturation density $\rho_0$. The calculation predicts a phase transition to pure quark matter at a density of 4 $\rho_0$ and at a temperature of about 50 MeV. This phase transition occurs shortly before the high-mass neutron star collapses into a black hole.

The question, whether baryons and quarks coexists at densities above 4-5 $\rho_0$, or whether already pure quark matter is created above 4 $\rho_0$ as suggested by figure 8, or whether still hadronic matter exists at this density, could be decided by heavy-ion collision experiments. As will be discussed in the following, observables like (i) multi-strange hyperons, (ii) event-by-event fluctuations of conserved quantities, and, finally, (iii) dileptons measured in heavy-ion collisions at different beam energies might help to shed light on the state of matter at high densities.

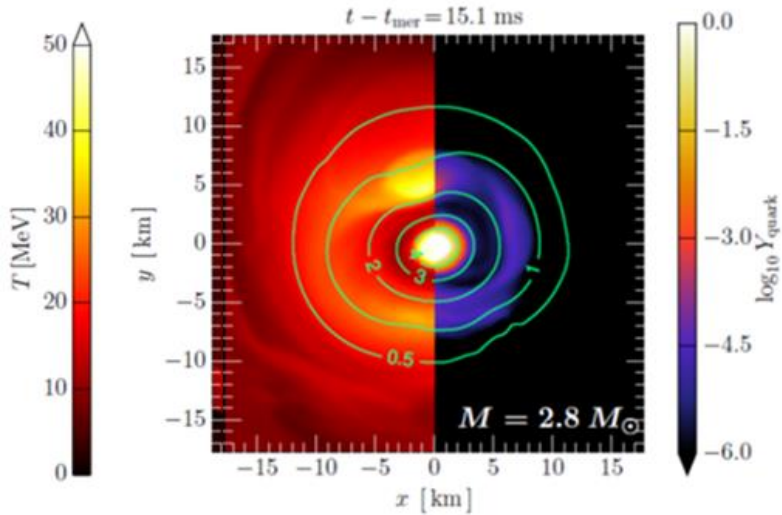

Fig.6: Snapshot of the evolution of two merging neutron stars with a total mass of 2.8 solar masses. The left and the right part of the plot indicates the temperature and the $\log_{10}$ of the quark fraction $Y_{quark}$, respectively. The green lines represent contours of constant net-baryon density in units of $\rho_0$. Adapted from [28].

*3.1 Probing the onset of deconfinement with multi-strange hyperons*

In heavy-ion collisions at ultra-relativistic beam energies it was found that the yield of multi-strange hyperons can be explained by statistical models, which assume a thermally equilibrated source. As the hyperon-nucleon scattering cross section is too small to drive the hyperons in equilibrium, it was claimed, that the equilibrium can only be achieved by multiple collisions at the phase boundary, i.e. during the phase transitions from the quark-gluon plasma to hadronic matter [29]. Also in heavy-ion collisions down to a beam energy of 30A GeV, the yield of multi-strange hyperons can be described by the statistical model [30]. However, in Ar + KCl collisions at an energy of 1.76A GeV, the yield of $\Xi^-$ hyperon measured by the HADES collaboration exceeds the statistical model prediction by a factor of 24±9, while all other measured hadrons are in agreement with the statistical model [31]. The research program of the CBM experiment at FAIR includes the measurement of the excitations function of hyperon production in heavy-ion collisions, in order to find the energy where multi-strange hyperons are driven into equilibrium, which would be an indication for the onset of deconfinement.

*3.2 Searching for the critical point with event-by-event fluctuations of conserved quantities*

The STAR experiment at RHIC has performed a beam energy scan searching for signatures of a first order phase transition. One of the most intriguing results was the event-by-event measurement of the 4th order cumulant (kurtosis) of the net-baryon distribution. This observable reflects density fluctuations in the reaction volume, which are similar to the phenomenon of critical opalescence, and, hence, are expected to occur in the vicinity of a critical point. This fluctuation signal starts to rise with decreasing beam energies below a collision energy of $\sqrt{s_{NN}} \approx 20$ GeV, and reaches its highest value at $\sqrt{s_{NN}} \approx 7.7$ GeV, which so far is the lowest energy accessible in the collider mode [32]. Already at this energy, the collision rate at RHIC has dropped to a few reactions per second, which prevents a further reduction of the collision energy. So the question is still open, whether, and if yes, at which beam energy this fluctuation signal exhibits a maximum, which then might be related to a critical point. Therefore, the FAIR beam energy range is well suited to search for the critical point, and the measurement of event-by-event fluctuation of conserved quantities is an important aspect of the CBM research program.

*3.3 Measuring the caloric curve with di-leptons*

The temperature of the fireball produced in heavy-ion collisions can be directly determined from the invariant mass spectra of di-leptons. The unique feature of an invariant mass spectrum is, that it is not blue-shifted, i.e. not affected by the collective expansion of the fireball, and, hence, its spectral slope reflects the fireball temperature if dilepton contributions from other sources are carefully subtracted. This has been recently demonstrated by the HADES collaboration, which measured at GSI the di-electron invariant mass spectrum for Au+Au collisions at 1.25A GeV. At this beam energy, densities of up to about 2.5 $\rho_0$ are expected to be created in the fireball. The HADES collaboration corrected the spectrum by known contributions from vector meson decays, which resulted in an exponential spectral slope reflecting a temperature of about 72 MeV [33]. This spectrum can be regarded as a direct measurement of the average temperature of the fireball, integrated over the collision history. Therefore, dilepton measurements open the unique possibility to determine the caloric curve of hot and dense QCD matter, which would be a direct experimental signature for a first order phase transition.

The CBM experiment at FAIR will continue the measurement of dilepton spectra towards higher beam energies, and will also extend the measured invariant mass range up to values of 2.5 GeV/$c^2$ or even higher. Dilepton invariant mass spectra between 1 – 2.5 GeV/$c^2$ are not contaminated by di-leptons from vector meson decays, and, hence, provide an even more direct information on the average fireball temperature [34].

*3.4 Combining observables towards a consistent picture*

Signatures for a QCD phase transition based on a single observation in heavy-ion collisions as described above always leave some room for alternative interpretations. However, the persuasive power of an explanation increases substantially, if results of several independent measurements can be explained consistently by a single scenario. Such a hypothetical situation is illustrated in figure 7, which depicts the emitting source temperatures as measured by HADES [33] and NA60 [35] as black symbols. The red dashed line illustrates the temperature calculated with a fireball model and a coarse-graining approach, extracted from the dilepton invariant mass range between 1–2 GeV/$c^2$ [36]. Such a curve is expected for a crossover transition, or for purely hadronic matter. The purple solid line represents a speculative caloric curve, indicating a first order phase transition from hadronic to quark-gluon matter. This curve starts from the onset of deconfinement, and ends in the critical point. The location of this critical point, indicated by the blue star symbol, is in agreement with the even-by-event fluctuation data of STAR. The future energy scan with the CBM detector will approach the critical point from lower beam energies, and will provide further constraints on its location. Finally, the planned measurement of the excitation function of multi-strange hyperon production is expected to indicate the onset of deconfinement, which may happen at an energy marked by the blue dot. In conclusion, if the different measurements would provide evidence for (i) the onset of deconfinement, (ii) for the critical point, and (iii) a caloric curve between these two points, one could claim the experimental discovery of a first order phase transition in dense baryonic matter.

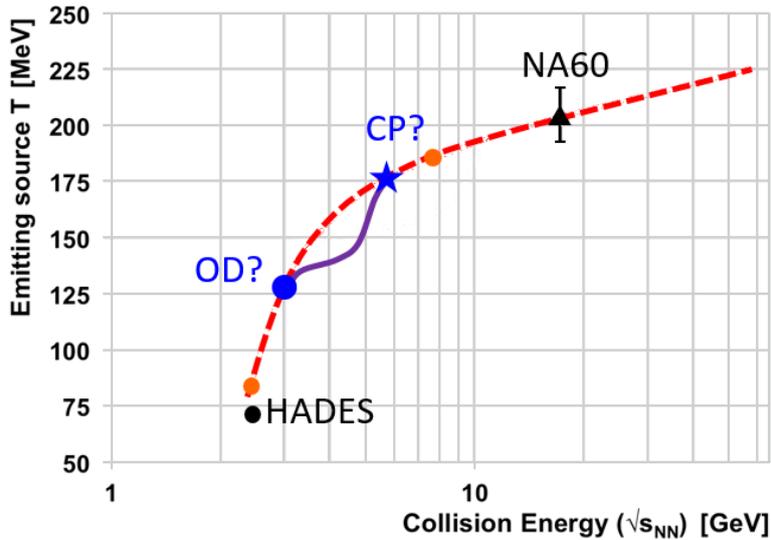

Fig.7: Temperature of the emitting source versus collision energy. The black symbols represent the temperature measurements of HADES [33] and NA60 [35]. The red dashed line depicts the temperature derived from the intermediate dilepton invariant mass range (1–2 GeV/$c^2$), calculated with a fireball model and a coarse-graining approach [36]. Such a curve reflects a smooth crossover transition, or no phase transition. The purple solid line illustrates a speculative caloric curve. The blue symbols indicate the hypothetical positions of a critical point (CP) and the onset of deconfinement (OD) (see text).

A comprehensive review of the theoretical concepts and the experimental approaches related to the investigation of high-density QCD matter can be found in the CBM Physics Book [37]. A recent update of the CBM physics program is presented in [38].

## 4. The CBM experiment at FAIR

The research program on dense QCD matter at FAIR will be performed by the experiments CBM and HADES. The two setups will be operated alternatively. The HADES detector, with its large polar angle acceptance ranging from 18 to 85 degrees, is well suited for reference measurements with proton beams and heavy-ion collision systems with moderate particle multiplicities, such as Ni+Ni or Ag+Ag at low SIS100 energies. With the HADES detector, electron-positron pairs and hadrons including multi-strange hyperons can be reconstructed.

Most of the relevant observables such as multi-strange (anti-) hyperons, hypernuclei and dileptons with a large invariant mass are extremely rarely produced in heavy-ion collisions at FAIR energies. In addition, important experimental information can only be obtained from multi-differential observables, such as collective flow of identified particles as function of momentum and collision centrality. Therefore, the Compressed Baryonic Matter (CBM) experiment at the Facility for Antiproton and Ion Research (FAIR) is designed to run at extremely high interaction rates of up to 10 MHz, which is orders of magnitude higher than the rate capabilities of existing and heavy-ion experiments under construction, as illustrated in figure 8 [38]. Both the **S**olenoidal **T**racker **A**t **R**HIC (STAR) at the Relativistic Heavy-Ion Collider (RHIC) in Brookhaven, and the future Multi-Purpose Detector (MPD) at the **N**uclotron-based **I**on **C**ollider f**A**cility (NICA) in Dubna are equipped with Time-Projection Chambers (TPC) as central tracking detectors, which are limited in their readout speed. The reaction rates of these experiments further decrease towards lower beam energies due to the decreasing luminosity of the colliders.

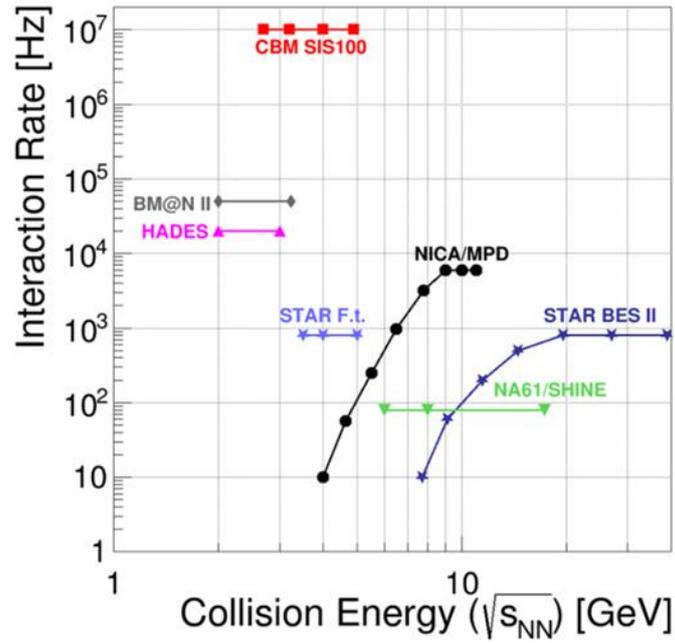

Fig. 8: Interaction rates achieved by existing and heavy-ion experiments under construction as function of collision energy (taken from [38]).

Figure 9 depicts the CBM detector system, which is designed to identify stable and short-lived hadrons, electrons and muons in nucleus-nucleus collisions. For example, when bombarding a 1 % interaction Au target with $10^9$ Au ions per second with a beam energy of 10A GeV, about $10^{11}$ charged particles have to be measured each second. Moreover, the tracks have to be reconstructed, and the particles and their decay products have to be identified.

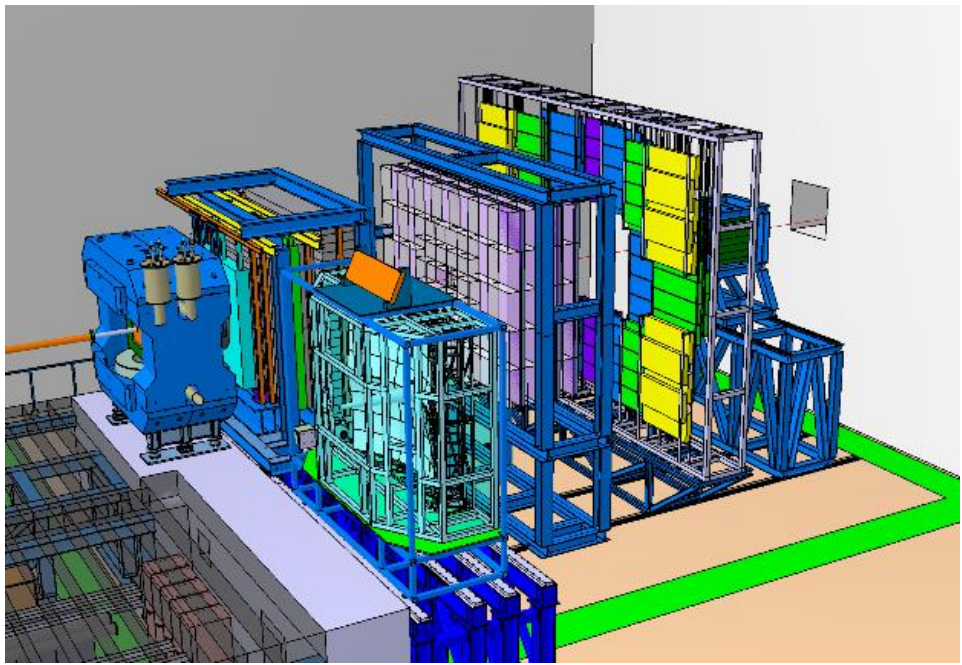

Fig.9: The CBM experimental setup (see text).

## 4.1 The CBM detector system

The superconducting dipole magnet hosts two tracking devices: A Micro-Vertex-Detector (MVD), consisting of four layers of silicon monolithic active pixel sensors located between 5 and 20 cm downstream the target, provides high-precision information of secondary decay vertices, e.g. of open charm mesons. The Silicon Tracking System (STS), comprising 8 layers of double-sided micro-strip sensors located between 30 and 100 cm downstream the target. The STS measures the tracks of charged particles, which is required for momentum determination. In order to identify the particles, their time-of-flight is measured with a wall of Multi-Gap Resistive Plate Chambers (MRPC), which has an active area of 120 $m^2$ at a distance of 7 m from the target. Electrons and positrons are separated from pions by two detectors: a Ring Imaging Cherenkov (RICH) detector and a Transition Radiation Detector (TRD). Four large-acceptance tracking detector stations, sandwiched between hadron absorbers made of graphite and iron blocks, provide muon identification. These detectors consist of Gas Electron Multiplier (GEM) detectors, and of Multi-Gap Resistive Plate Chambers. The muon chambers (MuCh) will be operated alternatively to the the RICH, and the TRD will be used for muon tracking after the last hadron absorber. The reaction plane angle of the collision will be measured with the Project Spectator Detector, which is a segmented hadronic calorimeter located about 10 m downstream the target.

In a heavy-ion collision with hundreds of tracks, the track topology of rare observables such as multi-strange hyperons or dileptons cannot be used to generate a hardware trigger. Therefore, a triggerless free-streaming data read-out and data acquisition system was developed for the CBM experiment. The front-end electronics generates a time stamp for each detector hit, which will be sent to the high-performance computing center of GSI ("GeenIT cube") without any event correlation. Here the tracks will be reconstructed, the particles will be identified, and finally the event will be defined and selected. This analysis is performed online, only based on the time and position information of the detector signals, by high-speed algorithms tuned to run on modern multi-core CPU architectures.

An actual overview of the status of the development of the CBM experiment with respect to detectors, electronics, data acquisition, software packages, and physic performance studies, can be found in the CBM Progress Report 2018 [39].

## 4.2 Physics performance studies

The layout of the CBM detector system has been performed by extensive simulations using event generators like UrQMD [40] and transport codes like GEANT3 [41]. The charged particle tracks were reconstructed using a Cellular Automaton algorithm [42]. Particle identification including short-lived particles like hyperons, hypernuclei, charmed particles etc., is performed by a newly developed software package based on the Kalman Filter approach [43], which takes into account also the time-of-flight information. In the following, some selected results on physics performance studies of planned hyperon and dilepton measurements will be presented.

In figure 10 the invariant mass spectra of $\Lambda$, $\Xi^-$, and $\Omega^-$ hyperons are presented, as simulated for central Au+Au collisions at 10A GeV using the UrQMD event generator, and reconstructed using the CBM software framework [44]. The signal-to-background ratio is about 4, 6, and 18 for $\Lambda$, $\Xi^-$, and $\Omega^-$, respectively, which is sufficient to perform the planned measurements. The resulting reconstruction efficiency for $\Xi^-$ hyperons, as function of transverse momentum $p_T$ and rapidity y, is shown in figure 11. The integrated $\Xi^-$ efficiency is about 20% above $p_T \approx 1$ GeV/c. Please note, that midrapidity is at y = 1.534, which is well covered by the acceptance of the detector system.

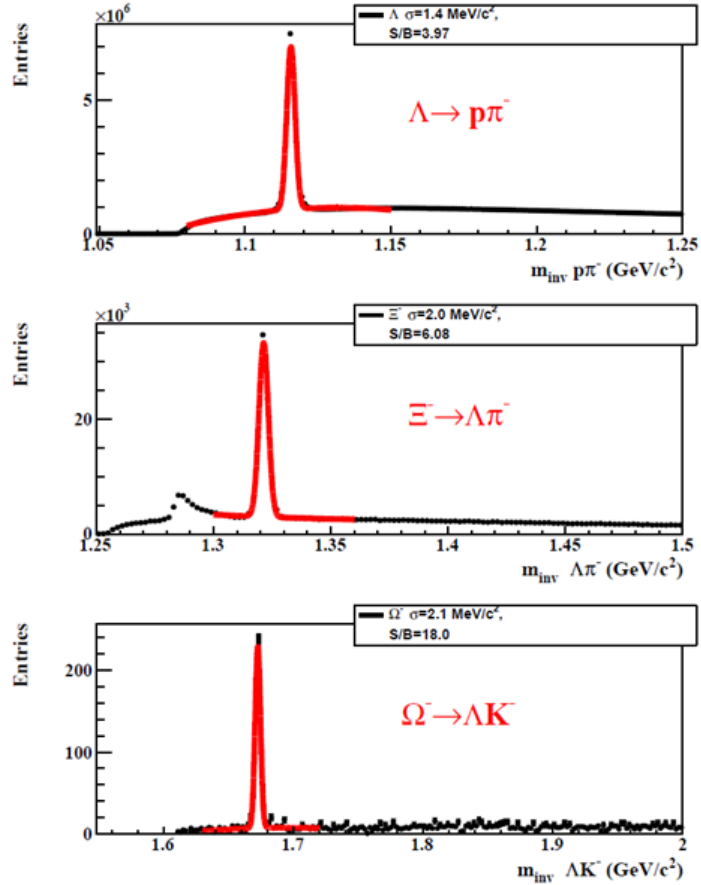

Fig.10: Invariant mass spectra of hyperons reconstructed from simulations of Au+Au collisions at 10A GeV based on a realistic description of the geometry and response of the CBM detectors. Taken from [44].

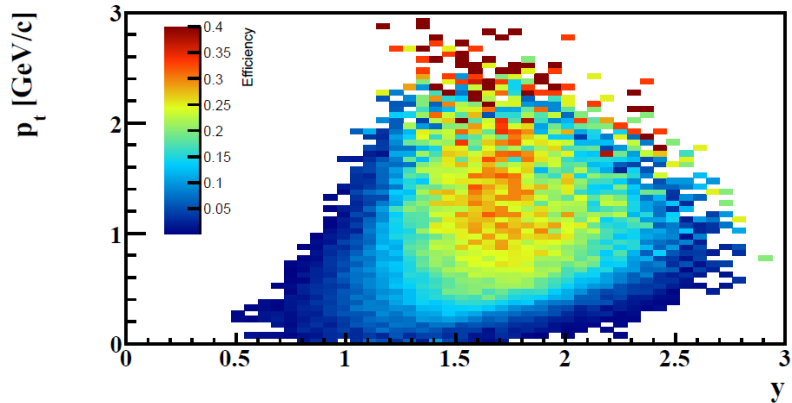

Fig. 11: Reconstruction efficiency for $\Xi^-$ hyperons as function of transverse momentum and rapidity, for in central Au+Au collisions at 10A GeV [44]. Midrapidity is at y = 1.534.

Figure 12 depicts the invariant mass spectrum of electron-positron pairs simulated with the UrQMD event generator for central Au+Au collisions at 8A GeV [45]. The reconstruction of the tracks and the identification of electrons and positrons relies on information from the MVD, STS, RICH, TRD and TOF detectors. The target thickness was 25 μm, in order to reduce the di-electron contribution from gamma conversion. The magnetic field was reduced to 50% of the maximum value. The contributions of vector meson decays, including their Dalitz decays, have been simulated with a thermal model [47]. The contribution from the in-medium ρ meson decay and the thermal radiation from the fireball (QGP) has been simulated using a coarse-graining approach [36].

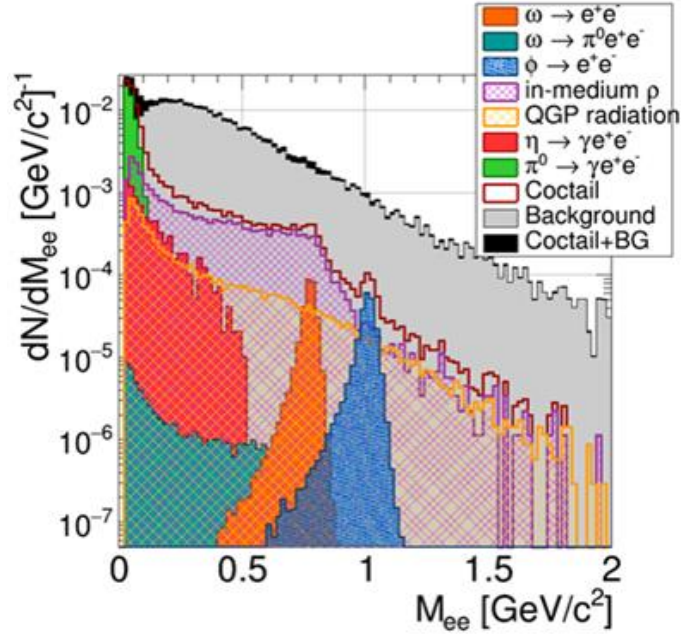

Fig. 12: Invariant mass spectrum of electron-positron pairs reconstructed from simulations of central Au+Au collisions at 8A GeV (see text).Taken from [45].

In figure 13 the result for the simulation of the dimuon invariant mass spectrum is shown, also for central Au+Au collisions at 8A GeV. In this case, the STS, MuCh, TRD and TOF detectors have been used. The MuCh consists of 4 tracking detector stations and 5 hadron absorber blocks, with the TDR as muon tracker after the last absorber [46]. Both the electron and muon simulations demonstrate that the CBM detector system is very well suited to measure and identify dilepton pairs in heavy-ion collisions at SIS100 energies.

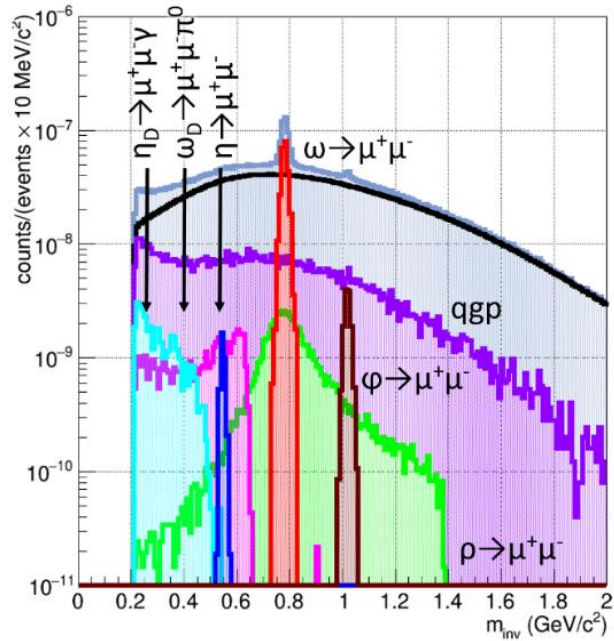

Fig.13: Invariant mass spectrum of muon pairs reconstructed from simulations of central Au+Au collisions at 8A GeV (see text).

## 5. Conclusions

The future FAIR accelerator center will offer unique possibilities to investigate fundamental questions related to astrophysics, including the origin of the elements in the universe, and the properties of QCD matter in compact stellar objects. While the superconducting Fragment-Separator and subsequent experimental setups will produce and study very rare neutron-rich or neutron- deficient isotopes playing a role in nucleosynthesis, the CBM experiment will explore the QCD phase diagram in the region of neutron star core densities. The CBM research program includes the exploration of the high-density equation-of-state, and the search for new phases of QCD matter. Promising observables sensitive to the EOS include the collective flow of identified particles, and the excitation function of multi-strange (anti-) hyperons. The measurement of event-by-event fluctuations of protons will probe the existence of the critical point, and eventually further constrain its location. The equilibration of multi-strange hyperons is expected to occur with the onset of deconfinement. The invariant mass spectra of dileptons directly reflect the temperature of the fireball. The discovery of a caloric curve, in combination with additional experimental evidence for the onset of deconfinement and for the critical point, would be a unique proof for a first order phase transition in dense QCD matter.

The successful execution of the CBM research program requires high-precision measurements of multi-differential distributions of the relevant diagnostic probes, in particular multi-strange particles, lepton pairs, and hypernuclei. The experimental challenge is to run at unprecedented heavy-ion reaction rates of up to 10 MHz, which demands the development of fast and radiation hard detectors, free streaming front-end electronics, and high-speed algorithms for online event reconstruction and selection running on a high-performance computing farm. Extensive physics feasibility studies of the envisaged hadron and dilepton measurements based on realistic detector geometries and response functions have been performed to optimize the experimental setup. First beams from the FAIR accelerators are expected to be delivered to the CBM experiment in 2025.

**Acknowledgment**

More than 470 scientists from 58 institutions and 12 countries work on the realization of the CBM experiment at FAIR, which is supported by GSI, the Helmholtz Association, the German Ministry of Education and Research, and by national funds of the CBM member institutions. The project receives funding from the Europeans Union's Horizon 2020 research and innovation programme under grant agreement No. 871072. The author acknowledges support from the Ministry of Science and Higher Education of the Russian Federation, grant N 3.3380.2017/4.6 and by the National Research Nuclear University MEPhI in the framework of the Russian Academic Excellence Project (contract No. 02.a03.21.0005, 27.08.2013).